\documentclass[aps,prl,english,twocolumn,superscriptaddress,showpacs,floats]{revtex4}
\usepackage{epsfig,epsf,amsmath,amsthm,amsfonts,
            graphics,graphicx,amssymb,babel}

\newcommand{\Dzero}{D\O\ }

\def\met{\mbox{${\hbox{$E$\kern-0.6em\lower-.1ex\hbox{/}}}_T$}} 

\newcommand{\Tesla}{\ensuremath{\mathrm{T}}}
\newcommand{\TeV}{\ensuremath{\mathrm{Te\kern -0.1em V}}}
\newcommand{\GeV}{\ensuremath{\mathrm{Ge\kern -0.1em V}}}
\newcommand{\MeV}{\ensuremath{\mathrm{Me\kern -0.1em V}}}
\newcommand{\GeVCSq}{\ensuremath{\GeV\!/c^2}}
\newcommand{\GeVC}{\ensuremath{\GeV\!/c}}

\newcommand{\ppbar} {p\bar{p}}
\newcommand{\ttbar} {t\bar{t}}

\newcommand{\fbinv} {fb$^{-1}$}
\newcommand {\abseta} {|\eta|}
\newcommand{\mo}{m_{0}}
\newcommand{\azero}{A_{0}}
\newcommand{\miz}{m_{1/2}}
\newcommand{\tb}{\tan\beta}
\newcommand{\nall}{\tilde{{\chi}}^{0}}
\newcommand{\call}{\tilde{{\chi}}^\pm}
\newcommand{\ntwo}{\tilde{{\chi}}^{0}_{2}}
\newcommand{\cone}{\tilde{{\chi}}^\pm_{1}}
\newcommand{\none}{\tilde{{\chi}}^{0}_{1}}
\newcommand{\ttt}{l_tl_tl_t}
\newcommand{\ttl}{l_tl_tl_l}
\newcommand{\tll}{l_tl_ll_l}
\newcommand{\ttT}{l_tl_tT}
\newcommand{\tlT}{l_tl_lT}

\addto{\captionsenglish}{

}

\begin{document}
\lefthyphenmin=2
\righthyphenmin=3

%
%
\title{
Search for Supersymmetry in $\ppbar$ Collisions at $\sqrt{s}=1.96$~$\TeV$
Using the Trilepton Signature for Chargino-Neutralino Production.

}
\affiliation{Institute of Physics, Academia Sinica, Taipei, Taiwan 11529, Republic of China} 
\affiliation{Argonne National Laboratory, Argonne, Illinois 60439} 
\affiliation{University of Athens, 157 71 Athens, Greece} 
\affiliation{Institut de Fisica d'Altes Energies, Universitat Autonoma de Barcelona, E-08193, Bellaterra (Barcelona), Spain} 
\affiliation{Baylor University, Waco, Texas  76798} 
\affiliation{Istituto Nazionale di Fisica Nucleare Bologna, $^t$University of Bologna, I-40127 Bologna, Italy} 
\affiliation{Brandeis University, Waltham, Massachusetts 02254} 
\affiliation{University of California, Davis, Davis, California  95616} 
\affiliation{University of California, Los Angeles, Los Angeles, California  90024} 
\affiliation{University of California, San Diego, La Jolla, California  92093} 
\affiliation{University of California, Santa Barbara, Santa Barbara, California 93106} 
\affiliation{Instituto de Fisica de Cantabria, CSIC-University of Cantabria, 39005 Santander, Spain} 
\affiliation{Carnegie Mellon University, Pittsburgh, PA  15213} 
\affiliation{Enrico Fermi Institute, University of Chicago, Chicago, Illinois 60637} 
\affiliation{Comenius University, 842 48 Bratislava, Slovakia; Institute of Experimental Physics, 040 01 Kosice, Slovakia} 
\affiliation{Joint Institute for Nuclear Research, RU-141980 Dubna, Russia} 
\affiliation{Duke University, Durham, North Carolina  27708} 
\affiliation{Fermi National Accelerator Laboratory, Batavia, Illinois 60510} 
\affiliation{University of Florida, Gainesville, Florida  32611} 
\affiliation{Laboratori Nazionali di Frascati, Istituto Nazionale di Fisica Nucleare, I-00044 Frascati, Italy} 
\affiliation{University of Geneva, CH-1211 Geneva 4, Switzerland} 
\affiliation{Glasgow University, Glasgow G12 8QQ, United Kingdom} 
\affiliation{Harvard University, Cambridge, Massachusetts 02138} 
\affiliation{Division of High Energy Physics, Department of Physics, University of Helsinki and Helsinki Institute of Physics, FIN-00014, Helsinki, Finland} 
\affiliation{University of Illinois, Urbana, Illinois 61801} 
\affiliation{The Johns Hopkins University, Baltimore, Maryland 21218} 
\affiliation{Institut f\"{u}r Experimentelle Kernphysik, Universit\"{a}t Karlsruhe, 76128 Karlsruhe, Germany} 
\affiliation{Center for High Energy Physics: Kyungpook National University, Daegu 702-701, Korea; Seoul National University, Seoul 151-742, Korea; Sungkyunkwan University, Suwon 440-746, Korea; Korea Institute of Science and Technology Information, Daejeon, 305-806, Korea; Chonnam National University, Gwangju, 500-757, Korea} 
\affiliation{Ernest Orlando Lawrence Berkeley National Laboratory, Berkeley, California 94720} 
\affiliation{University of Liverpool, Liverpool L69 7ZE, United Kingdom} 
\affiliation{University College London, London WC1E 6BT, United Kingdom} 
\affiliation{Centro de Investigaciones Energeticas Medioambientales y Tecnologicas, E-28040 Madrid, Spain} 
\affiliation{Massachusetts Institute of Technology, Cambridge, Massachusetts  02139} 
\affiliation{Institute of Particle Physics: McGill University, Montr\'{e}al, Canada H3A~2T8; and University of Toronto, Toronto, Canada M5S~1A7} 
\affiliation{University of Michigan, Ann Arbor, Michigan 48109} 
\affiliation{Michigan State University, East Lansing, Michigan  48824}
\affiliation{Institution for Theoretical and Experimental Physics, ITEP, Moscow 117259, Russia} 
\affiliation{University of New Mexico, Albuquerque, New Mexico 87131} 
\affiliation{Northwestern University, Evanston, Illinois  60208} 
\affiliation{The Ohio State University, Columbus, Ohio  43210} 
\affiliation{Okayama University, Okayama 700-8530, Japan} 
\affiliation{Osaka City University, Osaka 588, Japan} 
\affiliation{University of Oxford, Oxford OX1 3RH, United Kingdom} 
\affiliation{Istituto Nazionale di Fisica Nucleare, Sezione di Padova-Trento, $^u$University of Padova, I-35131 Padova, Italy} 
\affiliation{LPNHE, Universite Pierre et Marie Curie/IN2P3-CNRS, UMR7585, Paris, F-75252 France} 
\affiliation{University of Pennsylvania, Philadelphia, Pennsylvania 19104} 
\affiliation{Istituto Nazionale di Fisica Nucleare Pisa, $^q$University of Pisa, $^r$University of Siena and $^s$Scuola Normale Superiore, I-56127 Pisa, Italy} 
\affiliation{University of Pittsburgh, Pittsburgh, Pennsylvania 15260} 
\affiliation{Purdue University, West Lafayette, Indiana 47907} 
\affiliation{University of Rochester, Rochester, New York 14627} 
\affiliation{The Rockefeller University, New York, New York 10021} 

\affiliation{Istituto Nazionale di Fisica Nucleare, Sezione di Roma 1, $^v$Sapienza Universit\`{a} di Roma, I-00185 Roma, Italy} 

\affiliation{Rutgers University, Piscataway, New Jersey 08855} 
\affiliation{Texas A\&M University, College Station, Texas 77843} 
\affiliation{Istituto Nazionale di Fisica Nucleare Trieste/\ Udine, $^w$University of Trieste/\ Udine, Italy} 
\affiliation{University of Tsukuba, Tsukuba, Ibaraki 305, Japan} 
\affiliation{Tufts University, Medford, Massachusetts 02155} 
\affiliation{Waseda University, Tokyo 169, Japan} 
\affiliation{Wayne State University, Detroit, Michigan  48201} 
\affiliation{University of Wisconsin, Madison, Wisconsin 53706} 
\affiliation{Yale University, New Haven, Connecticut 06520} 
\author{T.~Aaltonen}
\affiliation{Division of High Energy Physics, Department of Physics, University of Helsinki and Helsinki Institute of Physics, FIN-00014, Helsinki, Finland}
\author{J.~Adelman}
\affiliation{Enrico Fermi Institute, University of Chicago, Chicago, Illinois 60637}
\author{T.~Akimoto}
\affiliation{University of Tsukuba, Tsukuba, Ibaraki 305, Japan}
\author{M.G.~Albrow}
\affiliation{Fermi National Accelerator Laboratory, Batavia, Illinois 60510}
\author{B.~\'{A}lvarez~Gonz\'{a}lez}
\affiliation{Instituto de Fisica de Cantabria, CSIC-University of Cantabria, 39005 Santander, Spain}
\author{S.~Amerio$^u$}
\affiliation{Istituto Nazionale di Fisica Nucleare, Sezione di Padova-Trento, $^u$University of Padova, I-35131 Padova, Italy} 

\author{D.~Amidei}
\affiliation{University of Michigan, Ann Arbor, Michigan 48109}
\author{A.~Anastassov}
\affiliation{Northwestern University, Evanston, Illinois  60208}
\author{A.~Annovi}
\affiliation{Laboratori Nazionali di Frascati, Istituto Nazionale di Fisica Nucleare, I-00044 Frascati, Italy}
\author{J.~Antos}
\affiliation{Comenius University, 842 48 Bratislava, Slovakia; Institute of Experimental Physics, 040 01 Kosice, Slovakia}
% ...

\collaboration{CDF Collaboration\footnote{With visitors from $^a$Universiteit Antwerpen, B-2610 Antwerp, Belgium, 
$^b$Chinese Academy of Sciences, Beijing 100864, China, 
$^c$University of Bristol, Bristol BS8 1TL, United Kingdom, 
$^d$University of California Irvine, Irvine, CA  92697, 
$^e$University of California Santa Cruz, Santa Cruz, CA  95064, 
$^f$Cornell University, Ithaca, NY  14853, 
$^g$University of Cyprus, Nicosia CY-1678, Cyprus, 
$^h$University College Dublin, Dublin 4, Ireland, 
$^i$University of Edinburgh, Edinburgh EH9 3JZ, United Kingdom, 
$^j$Universidad Iberoamericana, Mexico D.F., Mexico, 
$^k$University of Manchester, Manchester M13 9PL, England, 
$^l$Nagasaki Institute of Applied Science, Nagasaki, Japan, 
$^m$University de Oviedo, E-33007 Oviedo, Spain, 
$^n$Queen Mary, University of London, London, E1 4NS, England, 
$^o$Texas Tech University, Lubbock, TX  79409, 
$^p$IFIC(CSIC-Universitat de Valencia), 46071 Valencia, Spain,
$^x$Royal Society of Edinburgh/Scottish Executive Support Research Fellow,
$^y$Istituto Nazionale di Fisica Nucleare, Sezione di Cagliari, 09042 Monserrato (Cagliari), Italy 
}}
\noaffiliation

%
%
\begin{abstract}
We use the three lepton and missing energy ``trilepton" signature to search for chargino-neutralino production with $2.0$~fb$^{-1}$ of integrated luminosity collected by the CDF~II experiment at the Tevatron
$\ppbar$ collider.
We expect an excess of approximately 11 supersymmetric events for a choice of parameters of the mSUGRA model, but our observation of 7 events is consistent 
with 
the standard model expectation of 6.4 events.  We constrain the mSUGRA model of supersymmetry and rule out chargino masses up to 145~GeV/$c^2$ for a specific choice of parameters.

\end{abstract}
\maketitle
Supersymmetry posits the existence of boson (fermion)
``superpartners'' for standard model fermions (bosons)~\cite{susy}. 
This resolves the ``hierarchy problem''~\cite{hierarchy} in the standard model (SM) wherein an artificial
cancellation of large mass terms is required for the Higgs boson
mass to be at the electroweak scale. The lightest
supersymmetric particle (LSP), if stable and neutral, is an excellent dark matter candidate~\cite{dm}. 
However, since superpartners have not been observed at the same masses as the SM particles,
supersymmetry (SUSY) cannot be an exact symmetry. 
There are several models for breaking SUSY while retaining its advantages~\cite{susy_review}. A leading model is mSUGRA~\cite{msugra}, a grand unified theory (GUT) 
that incorporates gravity. Its five parameters, $\mo$, $\miz$, $\tb$, $\azero$, and the sign of $\mu$, defined at the GUT energy scale, 
determine the superpartner mass spectrum and coupling values at all scales.

Charginos ($\call$'s) and neutralinos ($\nall$'s) are mass eigenstates formed
by the mixture of gauginos and higgsinos, which are the fermionic superpartners of the gauge
and Higgs bosons~\cite{msugra}. At the Tevatron, associated production
of the lightest chargino with the next-to-lightest neutralino, $\ppbar\to\cone\ntwo+X$, may occur with a detectable rate.
Depending on the mSUGRA parameter values, the $\cone$ and the $\ntwo$ can decay as follows:
$\cone \to \ell^\pm \nu \none$ and $\ntwo \to \ell^+ \ell^- \none$,
where $\ell = e,\mu,\tau$ and $\none$ is the stable LSP. The neutrino and the LSP's are weakly interacting and escape undetected. This gives three leptons with 
large missing transverse energy  
as our experimental signature for $\cone \ntwo$ production.
Since jets are abundant at hadron colliders while leptons are rare, the trilepton signature is perhaps the best avenue for observing SUSY events.

Prior searches at the LEP $e^+e^-$ collider 
exclude chargino masses below $103.5$~GeV/$c^2$ with minimal assumptions~\cite{lepcone}. 
At the Tevatron, the CDF and \Dzero collaborations have searched for
$\cone \ntwo$ production 
with 1~\fbinv~of data~\cite{cdf1fb,d0trilep}, but
these trilepton searches have placed no further constraints on mSUGRA beyond those imposed by LEP experiments.
In this analysis of $2.0$~\fbinv~of CDF data, we are able to probe
mSUGRA beyond the LEP limits due to higher statistics and an improved technique.

The CDF~II detector~\cite{cdf} has cylindrical symmetry around the $\ppbar$ beam axis.  Our analysis is restricted to the central region of the detector defined by 
pseudorapidity $\abseta<1.1$~\cite{cdfcoord}.  The tracking system, used to measure the trajectory and momentum of charged particles, consists of multilayered silicon strip detectors and a drift chamber in
a 1.4~$\Tesla$ solenoidal magnetic field. Particle energies are measured with concentric electromagnetic (EM) and hadronic calorimeters.  Muon detectors consisting 
of wire chambers and scintillators are placed at the outer radial edge of the detector to allow for the absorption of most other particles in the intervening material.

A typical electron deposits most ($\sim$93\%) of its energy in the EM calorimeter producing an electromagnetic shower.
We identify an electron as a track whose energy deposit is consistent with its momentum ($E/p$ requirement). ``Tight'' electrons satisfy electromagnetic 
shower shape requirements; ``loose'' electrons
satisfy a weaker shower shape criterion and need not meet the $E/p$ requirement.
A ``tight muon'' is a track which leaves a minimum-ionizing energy deposit in the calorimeter and is detected by a muon detector.
``Loose muons'' are minimum-ionizing tracks outside the coverage of the muon detectors. 
We do not explicitly identify $\tau$-leptons. Instead, the electron and muon candidates 
can come from $\tau$ decays. 
In addition, we allow isolated tracks~\cite{isolation} as indicators of the hadronic decays of $\tau$-leptons to single charged particles.
Together, the $e, \mu$, and isolated track 
selection makes this analysis sensitive to $\sim$85\% of $\tau$ decays. The isolated track category also serves to accept poorly reconstructed $e$'s and $\mu$'s, 
albeit at the expense of a higher background. 

We require that the candidate leptons be isolated from 
 hadronic activity in the detector. Lepton and isolated track candidates consistent with photon conversions or cosmic rays are rejected.
The selected leptons have a small contamination from hadrons misidentified as leptons. These, along with leptons from semileptonic $b$, $c$ quark decays and residual 
photon conversions, are labeled as ``fake leptons.''

Neutrinos and the LSP's escape the detector, leading to significant missing transverse energy $\met$ in the event. $\met$ is defined as the magnitude of $\vec{\met} 
= -\Sigma_{i}E^{i}_{\rm T}\hat{n}_i$, where the unit vector $\hat{n}_i$ in the azimuthal plane points from the beam axis to the $i^{th}$ calorimeter tower. 
We correct $\met$ for the presence of candidate muons and isolated tracks.

Trilepton candidate events are collected with triggers that require
at least one tight electron (muon) with $E_{\rm T}>18$~GeV ($p_{\rm T} >18$~\GeVC) or two tight electrons (muons) with $E_{\rm T}>4$~GeV ($p_{\rm T} >4$~\GeVC).
An event must have at least two leptons ($e$'s or $\mu$'s); the third ``lepton'' can be an isolated track, with the sum of the lepton charges required to be $\pm 1$.
We define five trilepton channels: 
{\em  $\ttt$,
$\ttl$, $\tll$, $\ttT$,} and {\em  $\tlT$} where {\em  $l_t$, $l_l$, } and {\em  $T$ } refer
to a tight lepton, a loose lepton, and an isolated track, respectively. Table~\ref{tab:cats} shows the lepton energy 
thresholds for the five exclusive channels.
\begin{table}[htbp]\centering
 \caption{ The $E_{\rm T}$ ($p_{\rm T}$) thresholds for electrons (muons, isolated tracks) for the five channels. $l_t$=tight lepton, $l_l$=loose lepton, and $T$=isolated track (lepton=$e,\mu$). }
 \begin{tabular}{cccc}
\hline 
\hline
Channel& && $E_{\rm T}$ ($p_{\rm T}$) {\small GeV (GeV/c)} \\ \hline
$\ttt$ & &&  15, 5, 5 \\
$\ttl$ & &&  15, 5, 10 \\
$\tll$ & &&  20, 8, 5 (10 if $\mu$) \\
$\ttT$ & && 15, 5, 5 \\
$\tlT$ & &&  20, 8 (10 if $\mu$), 5 \\
\hline
\hline
 \end{tabular}
 \label{tab:cats}
\end{table}

Several SM processes can mimic the trilepton signature. The leptonic decays of $WZ$, $ZZ$, and $\ttbar$ 
can produce three or more leptons.
Dilepton processes, such as Drell-Yan or $WW$ accompanied by a bremsstrahlung photon conversion (``brem conversion''), a 
fake lepton, or an isolated track, are also a source of background. 
For channels with isolated tracks, $W$'s produced with jets result in significant background when one jet gives a fake lepton and another gives 
an isolated track.

To remove backgrounds containing an on-shell $Z$, we reject events when 
the invariant mass of either of the oppositely charged lepton-lepton or
lepton-track pairs is between 76 and 106~GeV/$c^2$.  
We reject events with a fourth lepton or track with $p_{\rm T} > 10$~GeV/$c$, principally to reduce the $ZZ$ background.
To suppress backgrounds from Drell-Yan and resonances such as J/$ \Psi $ and $ \Upsilon $, we
require invariant masses of both pairs to be $> 13 $~GeV/$ c^2 $
and at least one mass to be $ > 20 $~GeV/$ c^2 $. The Drell-Yan background is
further suppressed by requiring $\met>20$~GeV and the azimuthal angle 
between oppositely charged lepton-lepton (lepton-track) pairs to be less than
$2.9$ $(2.8)$~radians. To suppress the $\ttbar$ background, we allow no more than one jet with $E_{\rm T} > $15~GeV and $|\eta|<2.5$ in an event.

The mSUGRA parameters for our nominal signal point are $\mo=60$~GeV/$c^2$, $\miz=190$~GeV/$c^2$, $\tb=3$, $\azero=0$, and $\mu>0$. These parameter values are typical of the mSUGRA region in which this analysis is sensitive.
Simulated signal events are generated using the sparticle mass spectrum from {\sc isajet}~\cite{isajet}, followed by hard scattering in {\sc pythia}~\cite{pythia}. We use the {\sc madevent}~\cite{madevent} generator for the $WZ$ background simulation~\cite{generators}. The remaining background samples are generated using {\sc pythia}. Final stages of all signal and background simulation consist of hadronization using {\sc pythia} followed by CDF II detector simulation using {\sc geant3}~\cite{geant}. For all generators we use the {\sc cteq5l}~\cite{cteq} parton distribution functions (PDF).

\begin{table}[htbp] \centering
\caption{ The number of expected events from background sources and for the nominal mSUGRA point. The number of observed events in data is also shown.
$l_t$=tight lepton, $l_l$=loose lepton, and $T$=isolated track (lepton=$e,\mu$). The sums shown (rightmost column) are illustrative and not used in this analysis. }
\label{tab:bg}
\vspace{1mm}
\begin{tabular}{ccccccc|c} 
\hline
\hline
Channel&&            $\ttt$&           $\ttl$&          $\tll$&           $\ttT$&           $\tlT$&       $\Sigma$channels\\ 
\hline
Drell-Yan&&   0.05&          0.01&          0.0&           1.63&          1.32&        3.01       \\
Diboson&&        0.29&          0.20&          0.08&           0.61&          0.38&    1.56       \\
Top-pair&&  0.02&          0.01&         0.03&           0.22&          0.18&          0.46 \\
Fake lepton&& 0.12&          0.04&         0.03&           0.75&          0.41&        1.35   \\
\hline
Total &&0.49&          0.25&          0.14&          3.22&          2.28&       6.4\\
Uncertainty&& $\pm$0.09&  $\pm$0.04&   $\pm$0.03&      $\pm$0.72&     $\pm$0.63&  \\
\hline
Observed&&     1&             0&            0&              4&            2&         7\\
\hline
SUSY Signal&&  2.3&          1.6&          0.7&            4.4&          2.4&        11.4\\
\hline
\hline
\end{tabular}
\end{table}

\begin{figure}[htbp]
\begin{center}
\includegraphics*[width=0.5\textwidth]{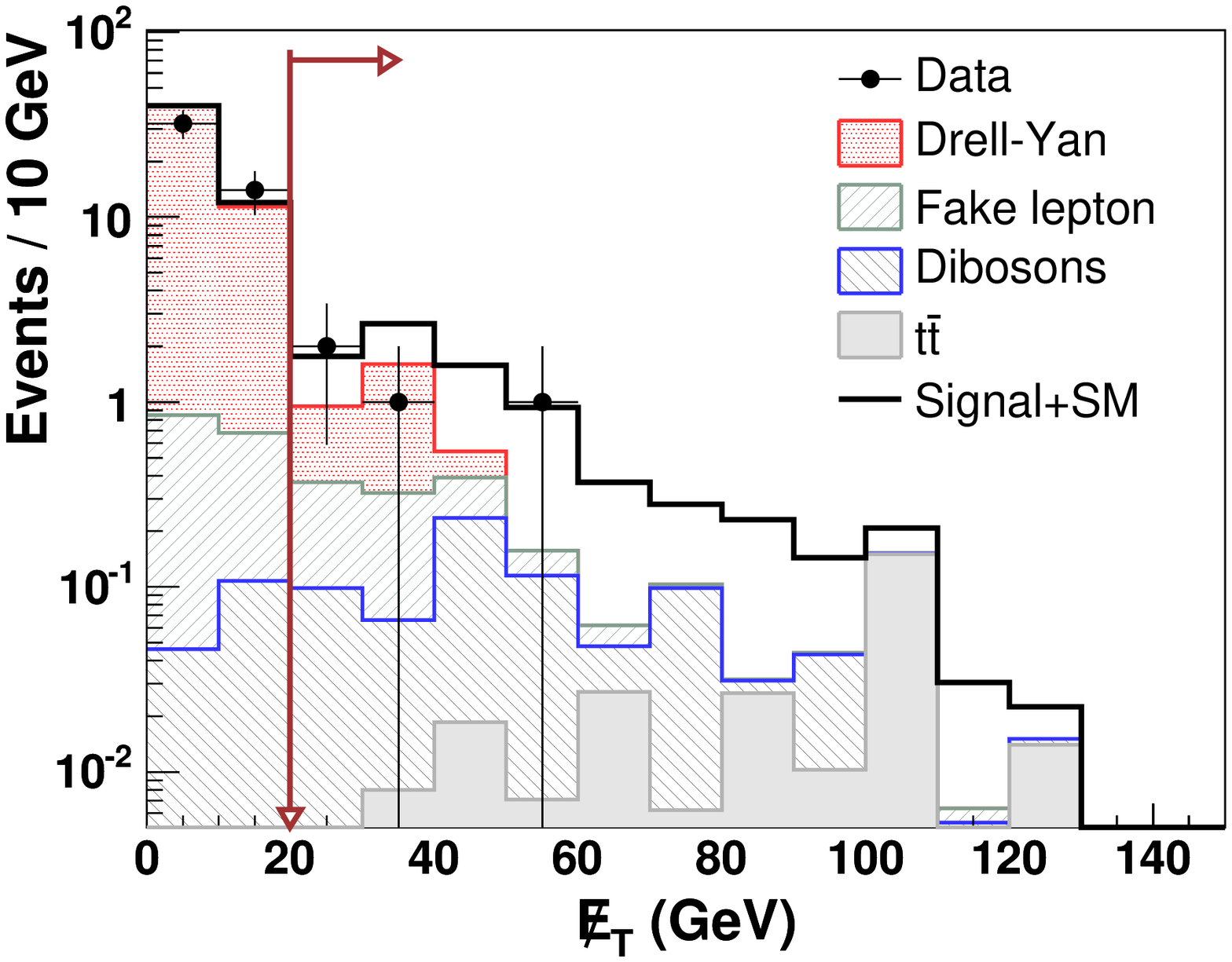}
   \caption{$\met$ distribution for the $\ttT$ 
channel after all other selection criteria are applied. 
The observed data (points) compare well to the stacked sum of standard model contributions.  The open histogram
shows the sum of SM contributions and expected signal for the nominal mSUGRA point. 
We keep events with $\met>$20~GeV. 
\label{fig:met}
}
\end{center}
\end{figure}

We estimate $WZ,ZZ$, Drell-Yan (+ brem conversion) and $WW$ (+ brem conversion) backgrounds using simulation. 
The number of events from Drell-Yan (or $WW$) + additional isolated track are determined by scaling the corresponding estimate from
simulation 
by the rate at which we expect an additional isolated 
track in the event. This rate is measured using $Z$ events in data.
We use data alone to measure backgrounds when a fake lepton accompanies a dilepton event by applying the fake lepton probability measured in the multijet data sample to the jets in the dilepton data events. We use the same technique when lepton+track processes such as $W+$jets result in background for channels with isolated tracks. Table~\ref{tab:bg} shows the number of expected background and signal events for the nominal signal point described above.

Table~\ref{tab:bg} also lists uncertainties due to various systematic sources. 
Significant uncertainties in signal (background) estimates are
4\% (2.5\%) due to lepton selection efficiency, 4\% (2.5\%) due to imperfect QCD radiation modeling, 2\% (1.5\%) due to the PDF uncertainty and 0.5\% (5\%) due to 
jet energy 
measurement uncertainty. 
There is a 6\% uncertainty in the integrated luminosity measurement and a 10\% theoretical uncertainty~\cite{plehn} in the signal cross section. 
The total background estimate has a 10\% uncertainty due to the lepton misidentification rate measurement, and uncertainties due to theoretical cross sections for 
background vary from 2.3 to 6.0\%.

\textit{Prior} to revealing candidate trilepton events in data, we verify the accuracy of background prediction for
numerous regions defined by $\met$ and invariant mass of the leading lepton pair. For example, in the 
region with  $\met > 15$~GeV and invariant mass from 76 to 106~GeV/$c^2$, i.e, a 15~GeV/$c^2$ window around the $Z$ mass, we predict $60.5\pm 9.1$ events and observe 61 trilepton events. 
A number of such ``control region'' comparisons and a detailed description of this analysis can be found in ~\cite{dube}. 
Figure~\ref{fig:met} shows the $\met$ distribution for the $\ttT$ 
channel. The observed and predicted distributions agree well over the entire range. The figure also shows the candidate trilepton events in this channel.
\begin{figure}[htbp]
  \begin{center}
    \includegraphics[width=0.5\textwidth]{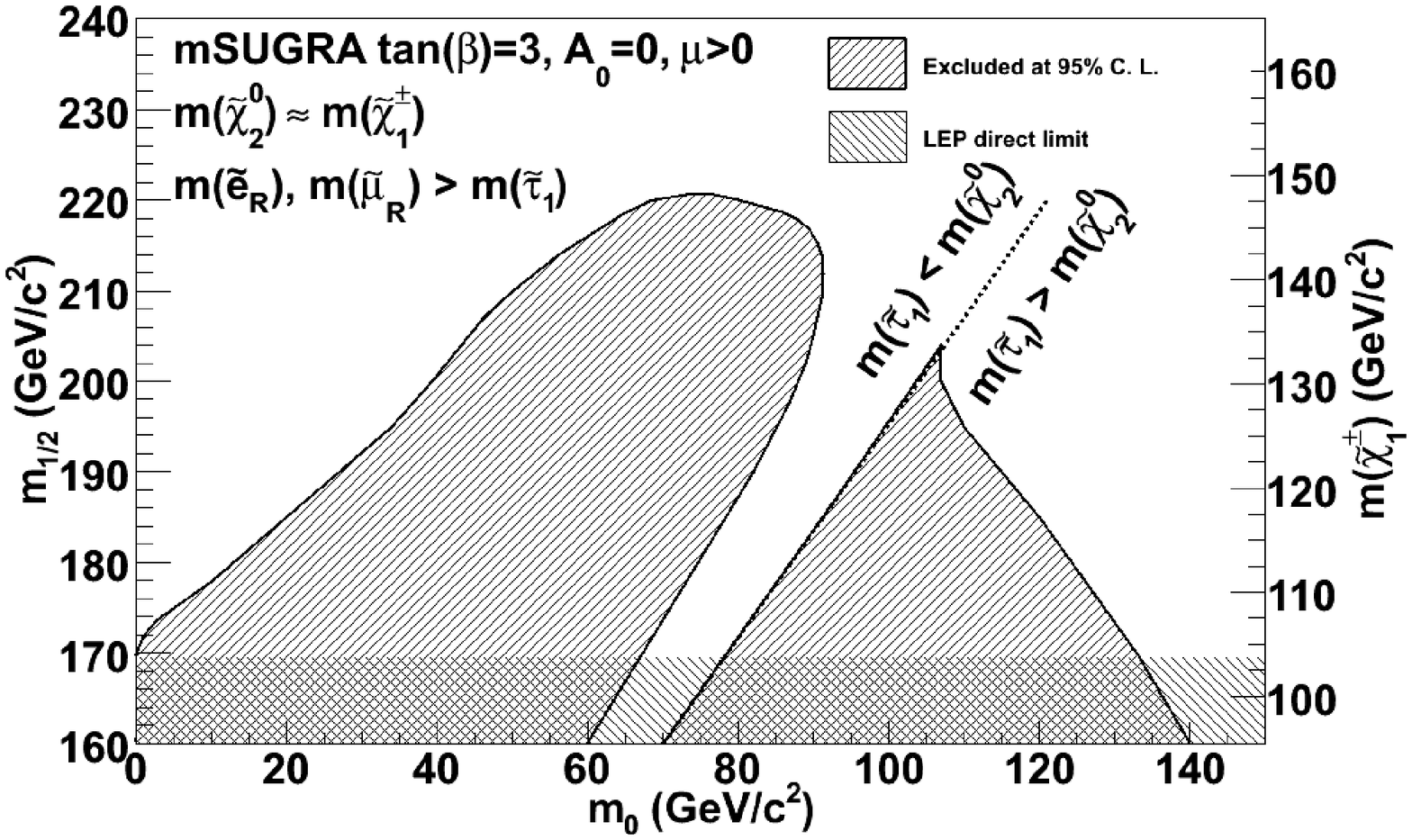}
    \caption{Excluded regions in mSUGRA as a function of $\mo$ and $\miz$ for $\tb=3$, $\azero=0$, $\mu>0$. Corresponding $\cone$ masses are shown on the right-hand axis.
\label{fig:exclusion} }
  \end{center}
\end{figure}


Table~\ref{tab:bg} shows that the observation is consistent with the expected background in each channel i.e. there is no evidence for 
physics beyond the standard model. 
We combine the individual channels following the method in \cite{junk,read} and calculate the limits on the cross section times branching fraction 
($\sigma\times{\cal B}$).  
Comparing the observed limits with the theoretically expected $\sigma\times{\cal B}$ calculated at next-to-leading order using {\sc prospino2}~\cite{prospino} gives the 95\% C.L. mSUGRA exclusion region. 

Figure~\ref{fig:exclusion} shows the exclusion in the mSUGRA $\mo$-$\miz$ plane divided in two regions: a heavy-slepton [m($\tilde{\tau}_1$)$>$m($\ntwo$)] region 
and a light-slepton [m($\tilde{\tau}_1$)$<$m($\ntwo$)] region.  In the heavy-slepton region, the $\cone$ and $\ntwo$ decay via virtual $W,Z$, or 
sleptons approximately equally to the three lepton flavors. In the light-slepton region the predominant decay of the $\cone$ and $\ntwo$ is via 
intermediate sleptons; the $\cone$ decays mostly via $\tilde{\tau}$, thus favoring a $\tau$-lepton in the final state. 
For small mass differences between $\ntwo$ and the sleptons (in Fig.~\ref{fig:exclusion}, the region immediately left of the dashed line), the decay of the $\ntwo$ 
via a slepton ($\ntwo\rightarrow \tilde{l}^\pm l^\mp$) leads to a lepton that is too soft to be detected. This loss in acceptance results in a gap in the exclusion 
between the two regions in the $\mo$-$\miz$ plane as seen in the figure.

\begin{figure}[htbp]
  \begin{center}
    \includegraphics[width=0.5\textwidth]{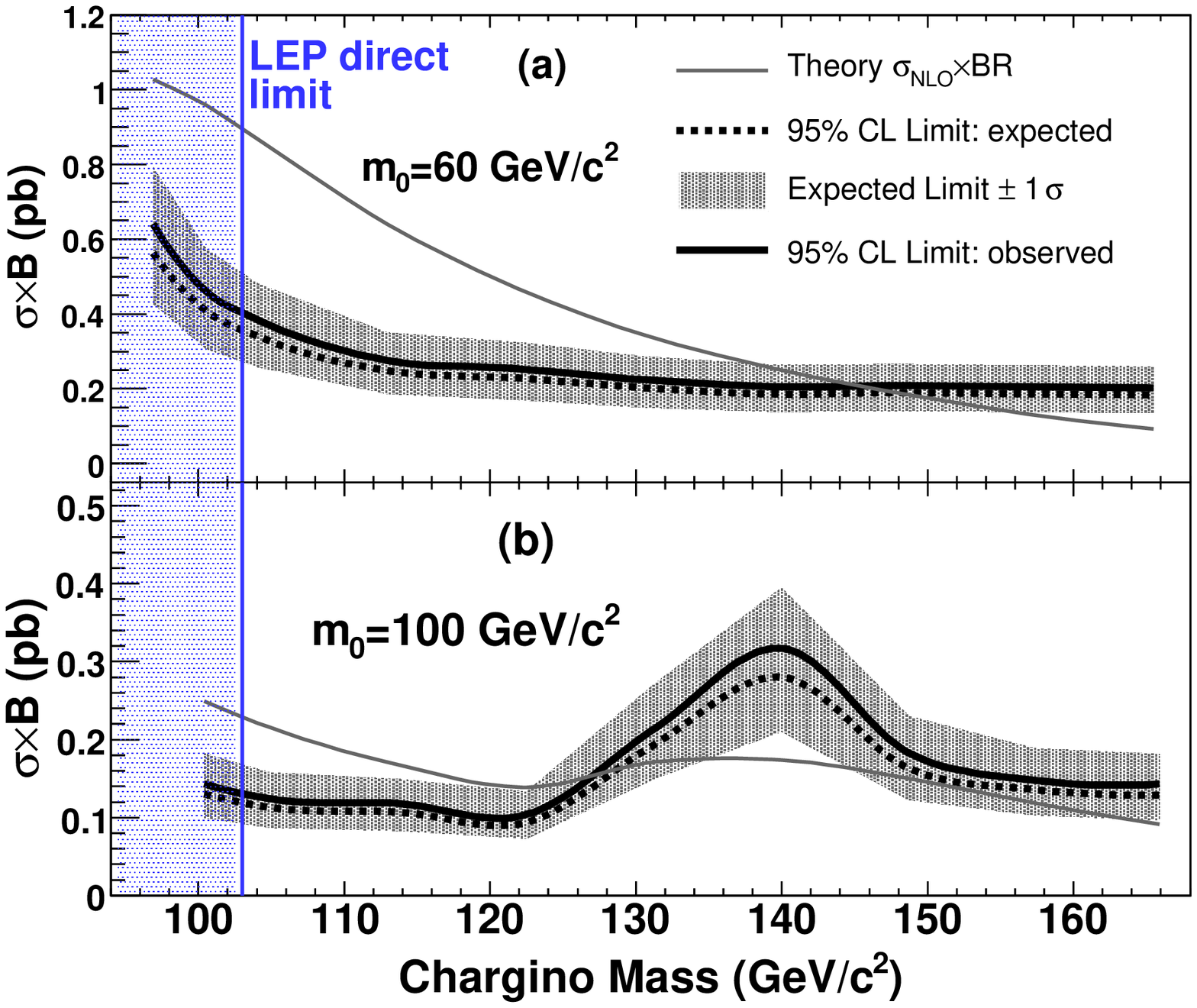}
    \caption{Observed and expected $\sigma\times{\cal B}$ and prediction from theory as a function of the chargino mass.
We rule out chargino masses below 145~\GeVCSq~(where the theory and the experimental
curves intersect) for (a)~$\mo=60$~\GeVCSq~ and below 127~\GeVCSq~ for (b)~$\mo=100$~\GeVCSq~. 
\label{fig:limit} }
  \end{center}
\end{figure}

We calculate the expected limits under the assumption that there is no physics beyond the SM.
Figure~\ref{fig:limit} compares the observed and expected $\sigma\times{\cal B}$ limits with the theoretical predictions for two choices of $\mo$: 60~GeV/$c^2$ 
in the light-slepton exclusion region and 100~GeV/$c^2$ in the heavy-slepton region. For the 60~GeV/$c^2$ case, we rule out chargino masses below 145~GeV/$c^2$.
For the 100~GeV/$c^2$  case, the limits worsen as the slepton becomes lighter than the chargino for chargino masses above
$\sim 130$~GeV/$c^2$. Once the softest lepton $p_T$ in the light-slepton region exceeds the detection threshold, the limit improves again. We rule out chargino masses below 127~GeV/$c^2$ in this case.

A study of how these trilepton search results apply to other SUSY models and to mSUGRA parameter space which we have not explored
here can be found in~\cite{pheno1}.

In conclusion, we have searched for chargino-neutralino production 
using the three lepton and large $\met$ signature with $2.0$~fb$^{-1}$ of data. 
Our observations are consistent with the standard model expectations. We 
exclude specific regions in the mSUGRA model's parameter space beyond LEP limits and
rule out chargino masses up to $145$~GeV/$c^2$ for a suitable choice of parameters.

We thank the Fermilab staff and the technical staffs of the participating institutions for their vital contributions. 
We thank S.~Thomas and M.~Strassler for their help with the theoretical interpretation.
This work was supported by the U.S. Department of Energy and National Science Foundation; the Italian Istituto Nazionale di Fisica Nucleare; the Ministry of Education, Culture, Sports, Science and Technology of Japan; the Natural Sciences and Engineering Research Council of Canada; the National Science Council of the Republic of China; the Swiss National Science Foundation; the A.P. Sloan Foundation; the Bundesministerium f\"ur Bildung und Forschung, Germany; the Korean Science and Engineering Foundation and the Korean Research Foundation; the Science and Technology Facilities Council and the Royal Society, UK; the Institut National de Physique Nucleaire et Physique des Particules/CNRS; the Russian Foundation for Basic Research; the Ministerio de Educaci\'{o}n y Ciencia and Programa Consolider-Ingenio 2010, Spain; the Slovak R\&D Agency; and the Academy of Finland.


\begin{thebibliography}{99}


\bibitem{susy} J.~Wess and B.~Zumino, Nucl. Phys. B~\textbf{70}, 39 (1974).

\bibitem{hierarchy}
E.~Witten, Nucl. Phys. B~\textbf{188}, 513 (1981);
N.~Sakai, Z. Phys. C~\textbf{11}, 153 (1981);
S.~Dimopoulos, Nucl. Phys. B~\textbf{193}, 150 (1981).

\bibitem{dm} J.~Ellis \textit{et al.}, Nucl. Phys. B~\textbf{238}, 453 (1984);
H.~Goldberg, Phys. Rev. Lett~\textbf{50}, 1419 (1983);  W.~M.~Yao \textit{et al.}, J. Phys. G~\textbf{33}, 1 (2006).
The stability of the LSP is a consequence of R-parity conservation.


\bibitem{susy_review} For a review, e.g., see H.~Haber and G.~Kane,
Phys. Rept. {\bf 117}, 75 (1985).

\bibitem{msugra} H.~P.~Nilles, Phys. Rep.~\textbf{110}, 1 (1984).

\bibitem{lepcone} LEP SUSY Working Group, LEPSUSYWG/01-03.1 (2001).

\bibitem{cdf1fb} T. Aaltonen \textit{et al.} (CDF Collaboration),
Phys. Rev. Lett.~\textbf{99}, 191806 (2007), and references therein. 

\bibitem{d0trilep} V.~Abazov \textit{et al.} (\Dzero Collaboration),
Phys. Rev. Lett.~\textbf{95}, 151805 (2005).

\bibitem{cdf} A. Abulencia \textit{et al.} (CDF Collaboration), J. Phys. G: Nucl. Part. Phys.~\textbf{34}, (2007).

\bibitem{cdfcoord} In the CDF cylindrical coordinate system,
$z$ axis is along the proton direction. Standard definitions are: 
$\theta$ = polar angle. Pseudorapidity $\eta= -\ln \tan(\theta/2)$.
Transverse momentum $p_{\rm T}=|p|\sin\theta$. Transverse energy
$E_{\rm T}= E\sin\theta$. 


\bibitem{isolation} For $e$'s and $\mu$'s, extraneous energy in the calorimeter in an $\eta$-$\phi$ cone of 0.4 around the lepton must be below 10\% of the lepton's energy. For tracks, the sum of momenta of other tracks within the cone and with $p_{\rm T}>0.4$~GeV/$c$ is restricted to less than 10\% of the track's $p_{\rm T}$. 



\bibitem{isajet} F.~Paige \textit{et al.}, hep-ph/0312045 (we use version 7.51 for simulated samples).


\bibitem{pythia}
T.~Sj\"{o}strand \textit{et al.}, Comput. Phys. Commmun. \textbf{135}, 238 (2001) (we use version 6.216).


\bibitem{madevent}
F.~Maltoni and T.~Stelzer, J.High Energy Phys.~\textbf{0302}, 027 (2003).


\bibitem{generators} The {\sc Madevent} generator is used for $\ppbar\rightarrow W\gamma^*$ in addition to  $\ppbar\rightarrow WZ$. The {\sc isajet} generator is used to calculate the mass spectrum of supersymmetric particles. {\sc pythia} is adequate for the remaining processes.


\bibitem{geant} R. Brun and F. Carminati, CERN Program Library Long Writeup W5013 (1993).

\bibitem{cteq} H.~L.~Lai \textit{et al.} (CTEQ Collaboration), Eur. Phys. J. C \textbf{12}, 375 (2000).


\bibitem{plehn} T.~Plehn and M.~Spira (private communication).

\bibitem{dube} S.~S.~Dube, ``Search for super symmetry at the Tevatron using the trilepton signature,'' FERMILAB-THESIS-2008-45.


\bibitem{junk} T.~Junk, Nucl. Instrum. Methods A~\textbf{434}, 435 (1999).

\bibitem{read} A. L. Read, J. Phys. G~\textbf{28}, 2693 (2002).



\bibitem{prospino} W.~Beenakker \textit{et al.}, Phys. Rev. Lett.~\textbf{83} 3780-3783, (1999) (Isajet v7.75 is used to calculate the sparticle spectrum input for {\sc prospino2}).

\bibitem{pheno1} S.~Dube, J.~Glatzer, S.~Somalwar, and A.~Sood, arXiv:0808.1605v1 [hep-ph].

\end{thebibliography}
\end{document}